\documentclass{article}

\usepackage{arxiv}

\usepackage[utf8]{inputenc} % allow utf-8 input
\usepackage[T1]{fontenc}    % use 8-bit T1 fonts
\usepackage{hyperref}       % hyperlinks
\usepackage{url}            % simple URL typesetting
\usepackage{booktabs}       % professional-quality tables
\usepackage{amsfonts}       % blackboard math symbols
\usepackage{nicefrac}       % compact symbols for 1/2, etc.
\usepackage{microtype}      % microtypography
\usepackage{graphicx}
\usepackage{doi}
\usepackage{blindtext}
\usepackage{enumitem}
\usepackage{epsfig}
\usepackage{graphicx}
\usepackage{amsmath}
\usepackage{amssymb}
\usepackage{multicol}
\usepackage{float}
\usepackage{mcite}
\usepackage[euler]{textgreek}
\usepackage{subcaption}
\usepackage{caption}
\usepackage{dblfloatfix}

\usepackage{physics}
\usepackage{siunitx}
\usepackage{cite}
\usepackage{braket}

\DeclareSIUnit\molar{\mole\per\cubic\deci\metre}
\DeclareSIUnit\Molar{\textsc{m}}

\setitemize[0]{leftmargin=*}
\setenumerate[0]{leftmargin=*}
\DeclareCaptionLabelFormat{cont}{#1~#2\alph{ContinuedFloat}}
\renewcommand{\headeright}{}
\renewcommand{\undertitle}{}
\renewcommand{\shorttitle}{The impact of a measurement on an open quantum system}

\hypersetup{
pdftitle={},
pdfsubject={},
pdfauthor={},
pdfkeywords={},
}

\begin{document}
\title{The impact of a measurement on an \\ open quantum system}

\author{C.J. Muller
\\ \\
This research has, in its entirety, been privately conducted and funded by the author.
	%% \AND
	%% Coauthor \\
	%% Affiliation \\
	%% Address \\
	%% \texttt{email} \\
	%% \And
	%% Coauthor \\
	%% Affiliation \\
	%% Address \\
	%% \texttt{email} \\
	%% \And
	%% Coauthor \\
	%% Affiliation \\
	%% Address \\
	%% \texttt{email} \\
}
\maketitle

\begin{abstract}
A molecular MCB junction in the partially wet phase has been used to probe effects related to open quantum systems. Although the exact quantum system, environment, and coupling, are not known the nature of the experiments shows a “measurement influenced” next measurement. The quantum system senses the measurement outcome and prepares itself in a state other than the state related to the measurement outcome. This triggers an alternation of measurements which are indicative of two current carrying states. In case three or more current carrying states are observed, there exists a fixed sequence of states that carry the current. We conclude that memory effects in these systems are responsible for these experimental observations.
\end{abstract}

\begin{multicols}{2}

% SECTION 1: Introduction
\section{Introduction}
Open quantum systems hold the key to our understanding how the quantum world transits to our classical world. In the early days of quantum mechanics, physicists were taught that the classical world was completely separated from the quantum world. Collapsing wave functions are at the interface between these two worlds, avoiding any difficulty explaining in detail how the successful quantum world, with its superposition principle and non-locality, evolved to our everyday world where we do not observe these effects. Concepts as “einselection”, “fragility of states” and “environment induced decoherence” \cite{zurec1}, \cite{zurec2}, \cite{Schlosshauer1} provide for a comprehensive basis where we continue to gain understanding on the fascinating quantum to classical transition. 

The Schr\"odinger equation describes quantum systems that are perfectly isolated from the environment and therefore remain coherent forever. To be able to perform a measurement on a quantum system we need to connect it to the environment \cite{breuer}. This connection is crucial and has initiated a vast amount of research and literature on open quantum systems. A quantum system connected to an environment starts to lose coherence, thus becoming more decoherent. Decoherence of a quantum system implies that some of its information is lost to the environment. The environment acts not only as a sink of coherence, but also as the transporter of information.

The connection between the quantum system and the environment can be unidirectional, allowing for a flow of information from the quantum system to the environment only, or it can be bidirectional. Unidirectional couplings are less likely to represent the actual physics of the system, however they can provide insight into open quantum systems in general. The quest related to open quantum systems is to reveal the master equation. The master equation provides the solution to how open quantum systems will evolve over time. The stage of such a system at $t=t_0$ is provided by system($t_0$). The future of this system is defined by an operator acting on system($t_0$). Thus, the future of the system is based solely on the stage of the system at $t_0$ and does not depend on the system prior to $t_0$. These systems are called Markovian, or memoryless. The Markovian master equation relies on probability conservation. 

For a bidirectional connection of the quantum system to the environment, physics becomes more interesting. We now have information flowing from the quantum system to the environment and this information may find its way back to the quantum system. There is a characteristic time involved in the flow of information from the quantum system to the environment. One can easily see that now memory effects start to play a role in the time evolution of this system. Time before $t_0$ will be required in the master equation describing the future of these systems. The difficulty is how much time and how large an environment must be included in the equations. In principle the environment can be the entire universe. The master equation for these systems contains a part which relates to the past of the system, this part contains the so called “memory kernel” which is integrated over a certain amount of time before $t_0$. These systems are called non-Markovian or systems with a memory. State tomography or process tomography are tools to reconstruct the time evolution of the system. In general modelling the time evolution of an open quantum system is much more difficult as compared to solutions of the Schr\"odinger equation for isolated quantum systems \cite{Pollock}, \cite{ModiK}. 

Here we present data which show that non-Markovianity plays a key role in the description of an open quantum system. The impact of a measurement on the above-described system is astonishing.

% SECTION 2
\section{Experimental setup and results}
The experimental set up and operation of a molecular MCB junction has been described in detail in reference \cite{Muller}. Experiments are performed at ambient conditions, the gold electrodes of the MCB junction are used to create the system: electrode-molecule-electrode. The entire electrode-molecule-electrode configuration is lined with a microscopic layer of tetrahydrofuran (THF), or equivalently, is in the partially wet phase. We present five current voltage (IV) curves, IV1-IV5 in corresponding figure numbers, from the bending beam assembly BBA1, recorded after each other. 
These IV curves provide the full cycle from start to end of BBA1 in the partially wet phase. The IV curves contain 1000 points per scan, a scan is recorded in 45 seconds. An IV curve contains two scans, one with an increasing voltage from \SI{9}{\volt} to \SI{10}{\volt} and one with decreasing voltage from \SI{10}{\volt} to \SI{9}{\volt} where there is no delay at the point of scan-reversal.

If partial traces are shown the time between measurement points remains \SI{45}{\milli\second}. The time between consecutive IV 

\begin{figure}[H]
    \centering
    \includegraphics[width=.5\textwidth]{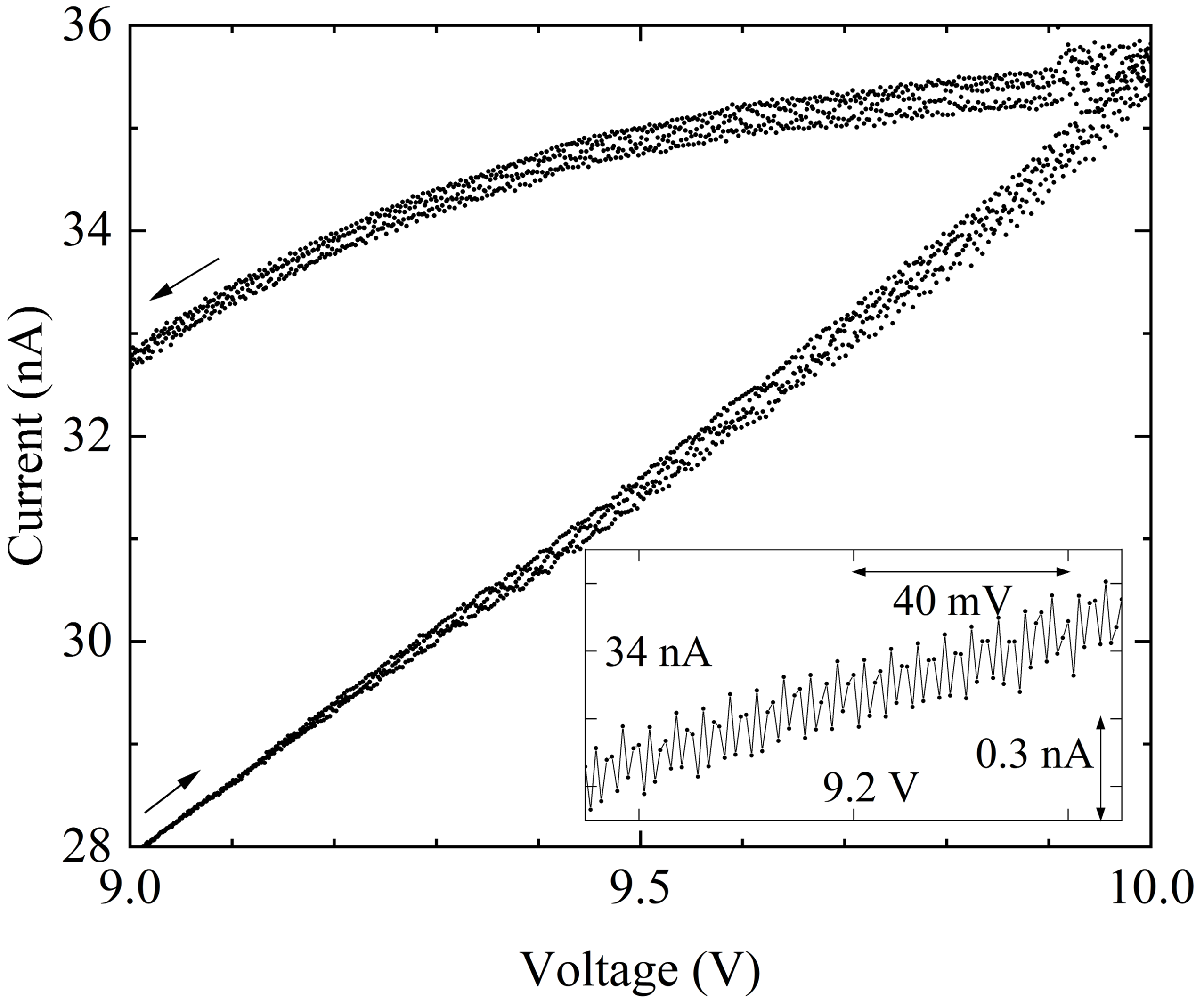}
    \caption{An IV curve showing a broad range of data containing some visible structure in the upper and lower IV curve. The inset reveals that this data follow a regular pattern.}
    \label{fig:IV01}
\end{figure}

curves is approximately 45 seconds.
Furthermore, data from two additional bending beam assemblies (BBA2 and BBA3) are presented to indicate that the reported results reproduce over various bending beam assemblies. The fluid used in the cell is a benzene dithiol (BDT)/THF solution, 1mMol/L for BBA1. A pure THF only solution has been used for BBA2 and BBA3. A THF molecule is bridging the gap of the gold electrodes for BBA2 and BBA3. For BBA1 where the BDT/THF solution is used, we cannot exclude the possibility that a THF molecule is the bridging molecule, thus either a BDT or a THF molecule will bridge the electrode gap.

The experiment starts with breaking the gold filament of the bending beam assembly in solution, creating two fresh gold electrodes. Draining the cell containing the liquid and letting the cell and electrodes dry will at a certain point end up in a single molecule captured between the electrodes where the electrodes and molecule will be lined with a THF partially wet phase layer. All recordings have been performed in the partially wet phase which endures approximately for up to ten minutes. For one bending beam assembly, BBA1, the five recorded IV curves provide an overview of the difference and wealth of effects that are measured.

The most important data are presented in fig. \ref{fig:IV02} and \ref{fig:IV03a}a. A remarkable pattern of measurement points is obtained, where a clear grouping of oscillating points is present. For example, at \SI{9.2}{\volt} in both the upper and lower IV curve, or from \SIrange{9.6}{10}{\volt} in the lower IV curve from fig. \ref{fig:IV02}. We will call such a group of data points which belong together a “state-curve.” A number of state-curves within a certain voltage range represents the same number of current carrying states within that range. It is anticipated that the one dimensional density of states peak at the Fermi level  probes these states \cite{Muller}.

\begin{figure}[H]
    \centering
    \includegraphics[width=.5\textwidth]{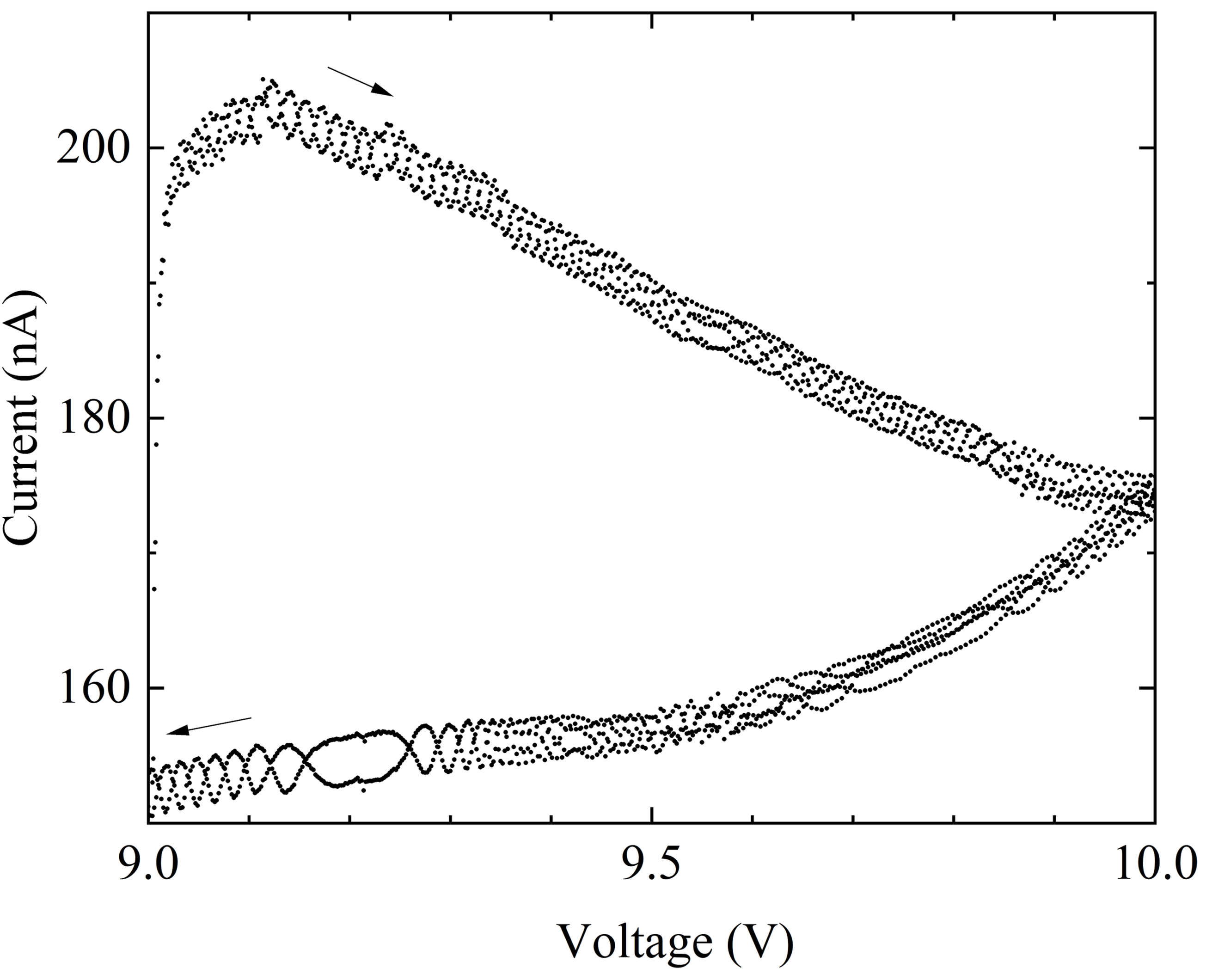}
    \caption{A spectrographic chart, where a number of current carrying states of the molecule are mapped is shown. Oscillating lines (state-curves) are displayed, intersecting one another in a certain pattern.}
    \label{fig:IV02}
\end{figure}

\begin{figure}[H]
    \ContinuedFloat*
    \centering
    \includegraphics[width=.5\textwidth]{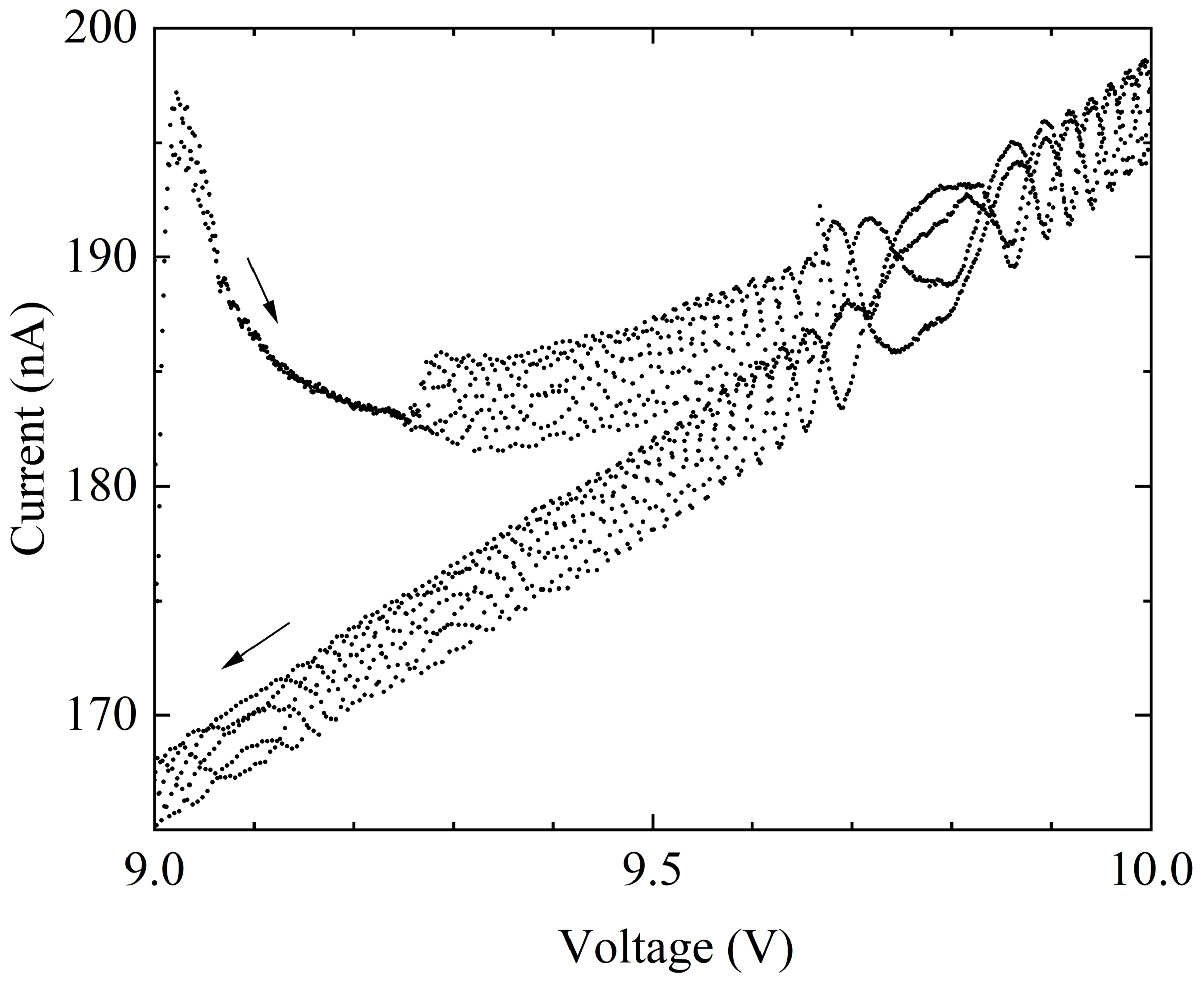}
    \caption{A next IV curve shows a similar pattern as compared to fig. \ref{fig:IV02} confirming the effect.}
    \label{fig:IV03a}
\end{figure}

\begin{figure}[H]
    \ContinuedFloat
    \centering
         \includegraphics[width=.5\textwidth]{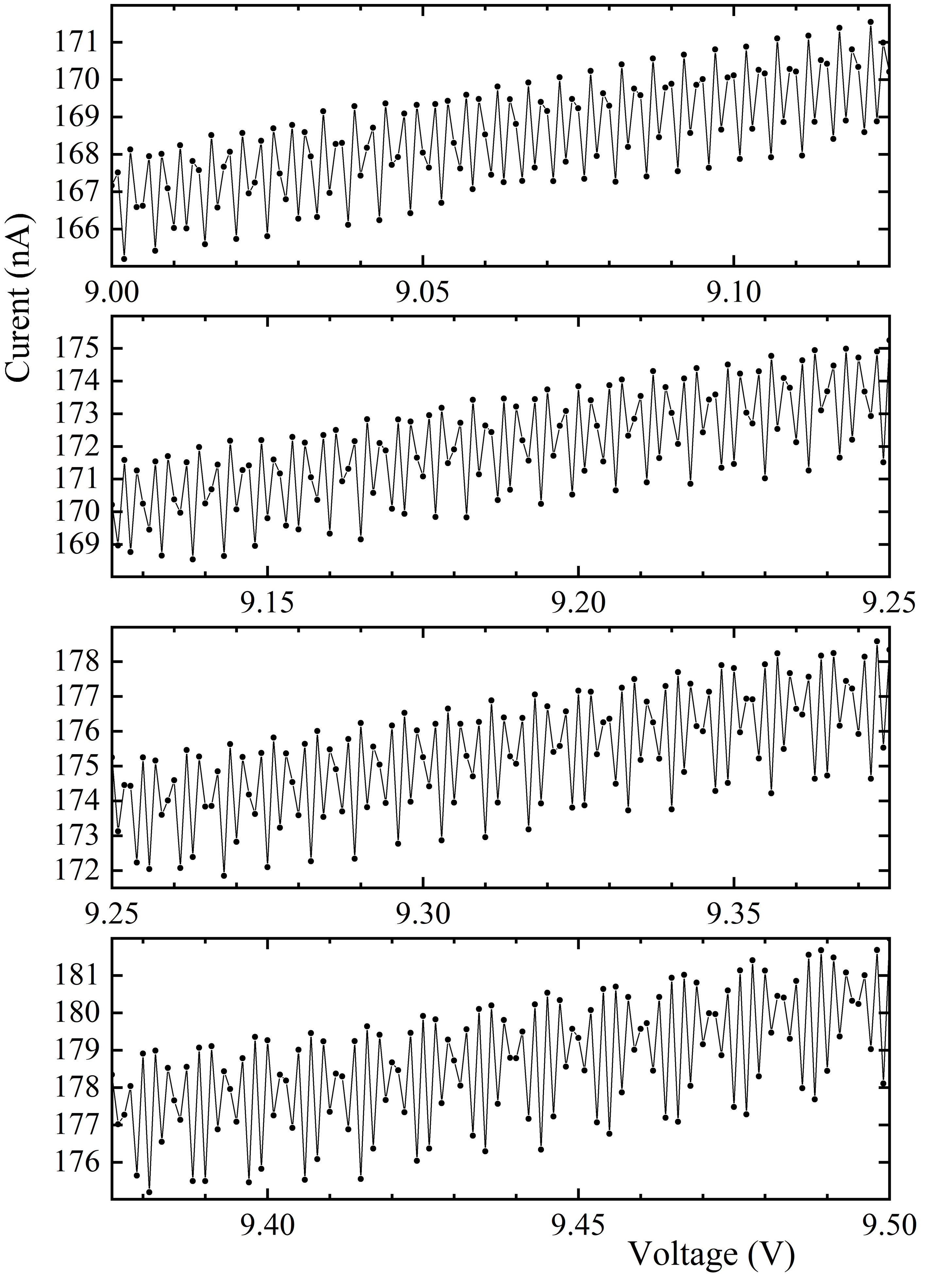}
         \caption{Additional state-curves originate from the data points at or near the nodes. The state-curves are clearly visible in fig. \ref{fig:IV03a}a, this figure shows the sequence of the data on an enlarged scale. }
         \label{fig:IV03b}
\end{figure}

A shifting pattern over the entire 2000 measurement point range is observed in fig. \ref{fig:IV02}. This pattern is confirmed by the IV curve in fig. \ref{fig:IV03a}a, showing a similar but different pattern. Also, in this case a pattern of state-curves builds. In fig. \ref{fig:IV03a}a, at a certain voltage value, two or more state-curves can be observed. For example, two state-curves at \SI{9.8}{\volt} (upper and lower IV curve), or five state-curves near \SI{9.1}{\volt} (lower IV curve), or seven state-curves near \SI{9.3}{\volt} (lower IV curve). In fig. \ref{fig:IV03b}b and fig. \ref{fig:IV03c}c the lower IV curve of fig. \ref{fig:IV03a}a is shown in detail. From the top two panels in fig. \ref{fig:IV03b}b it is obvious that the data generate a regular pattern. The state-curves in fig. \ref{fig:IV03a}a, especially in the lower IV curve from \SI{9}{V} to \SI{9.25}{V} are not immediately visible in the enlarged view in fig. \ref{fig:IV03b}b. This enlarged view of the data does however show the exact sequence of the data, which is difficult to see from fig. \ref{fig:IV03a}a. We need both an extended view of the data as well as the enlarged view to gain insight into the data. Fig. \ref{fig:BBA2} and \ref{fig:BBA3} show data similar in nature as shown in fig. \ref{fig:IV02} and \ref{fig:IV03a}a from two different bending beam assemblies. The inset of fig. \ref{fig:BBA2} and \ref{fig:BBA3} provides the full IV curve over the \SIrange{9}{10}{V} range. Clear state-curves are observed as a function of voltage, four state-curves in fig. \ref{fig:BBA2} and three state-curves
in fig. \ref{fig:BBA3} of which two state-curves seem to interact in the bottom IV curve. For fig. \ref{fig:BBA2} purple,

\begin{figure}[H]
    \ContinuedFloat
    \centering
    \includegraphics[width=.5\textwidth]{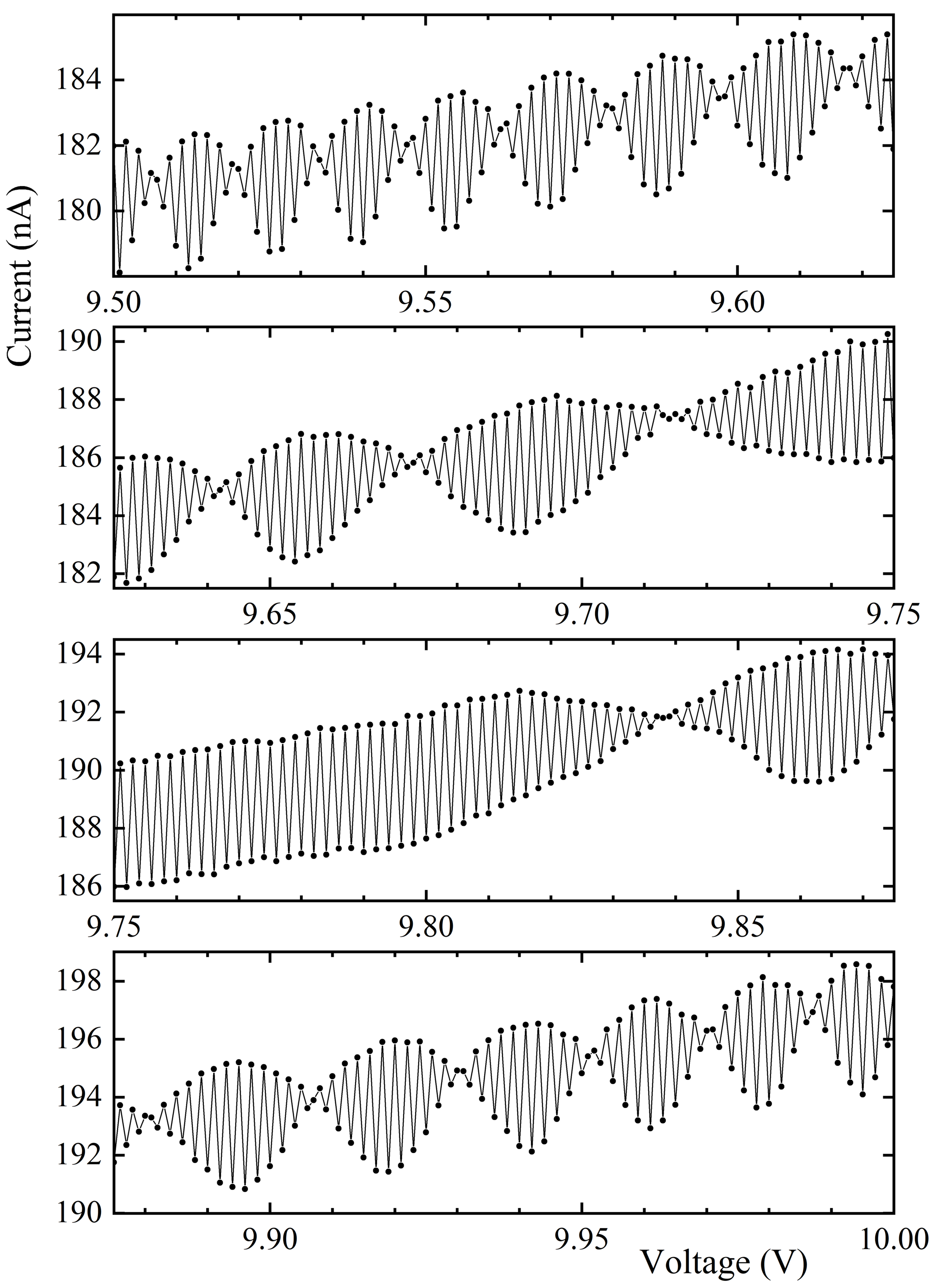}
    \caption{The min-max current bandwidth oscillates as a function of voltage. Well defined minima, called nodes, are observed in these panels.}
    \label{fig:IV03c}
\end{figure}

\pagebreak

 green, blue, and red points, belong to one corresponding state-curve all points advance the previous color by \SI{1}{\milli\volt} in this specific order for the entire voltage range shown.

\begin{figure}[H]
    \centering
    \includegraphics[width=.46\textwidth]{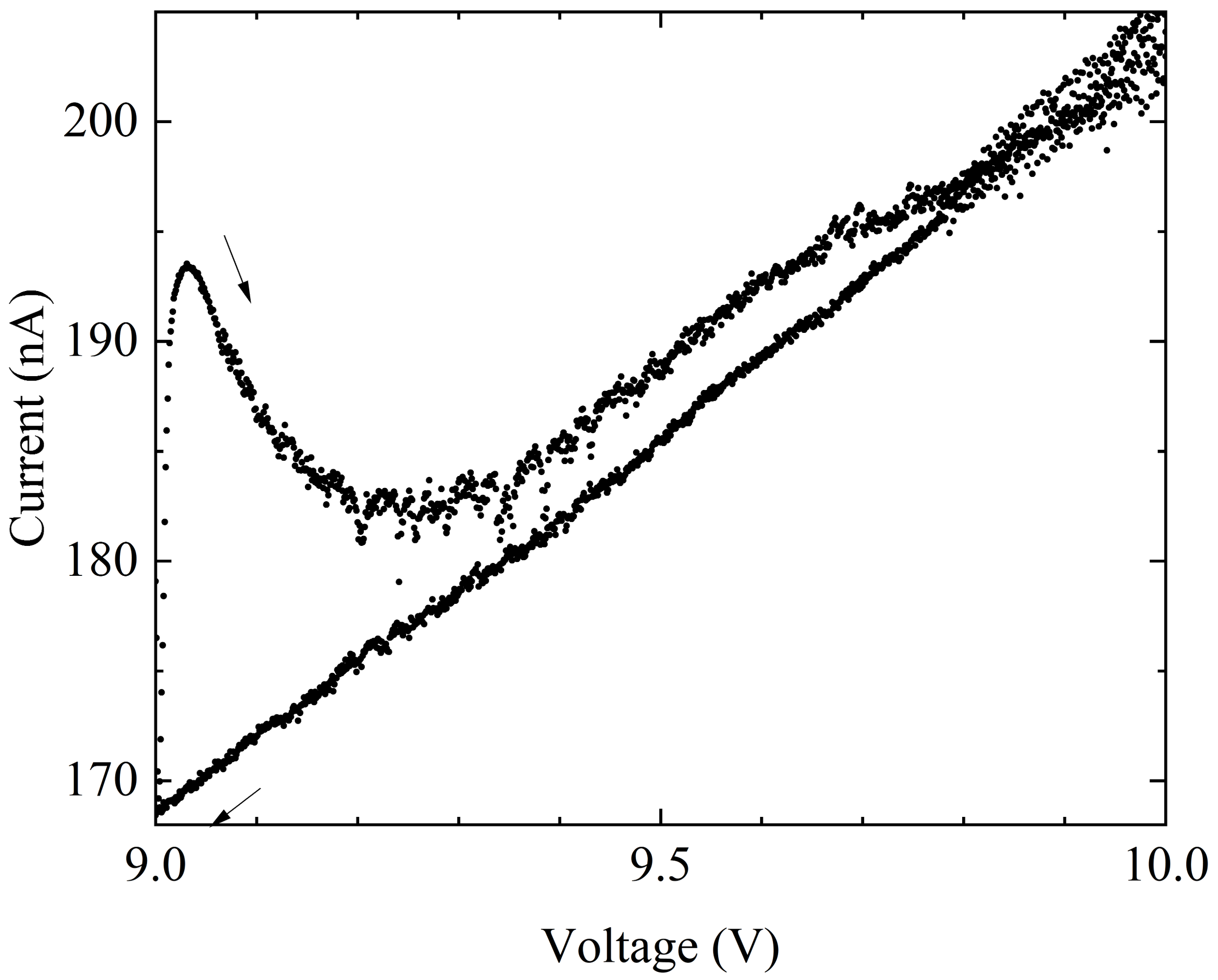}
    \caption{In this IV curve the regular pattern from the previous two IV curves has been transited to a noisier IV curve. }
    \label{fig:IV04}
\end{figure}

\begin{figure}[H]
    \centering
    \includegraphics[width=.47\textwidth]{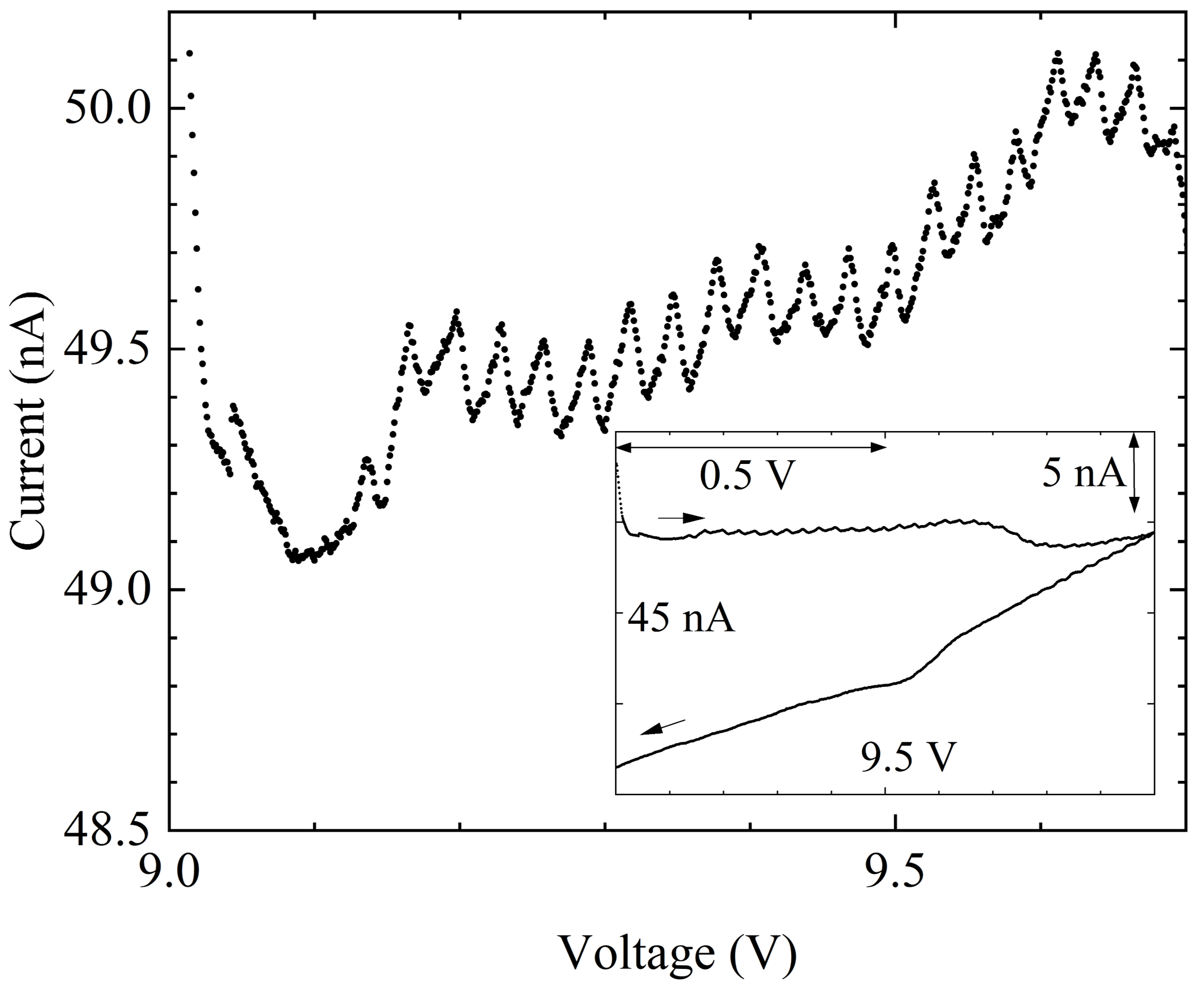}
    \caption{A considerable reduced average current in this IV curve as compared to the prior three IV curves is indicative that the completely dry phase is nearing. Clear sharp pointed quantum structure is visible. The inset shows the IV curve over the entire \SIrange{9}{10}{V} range. }
    \label{fig:IV05}
\end{figure}

\begin{figure}[H]
    \centering
    \includegraphics[width=.49\textwidth]{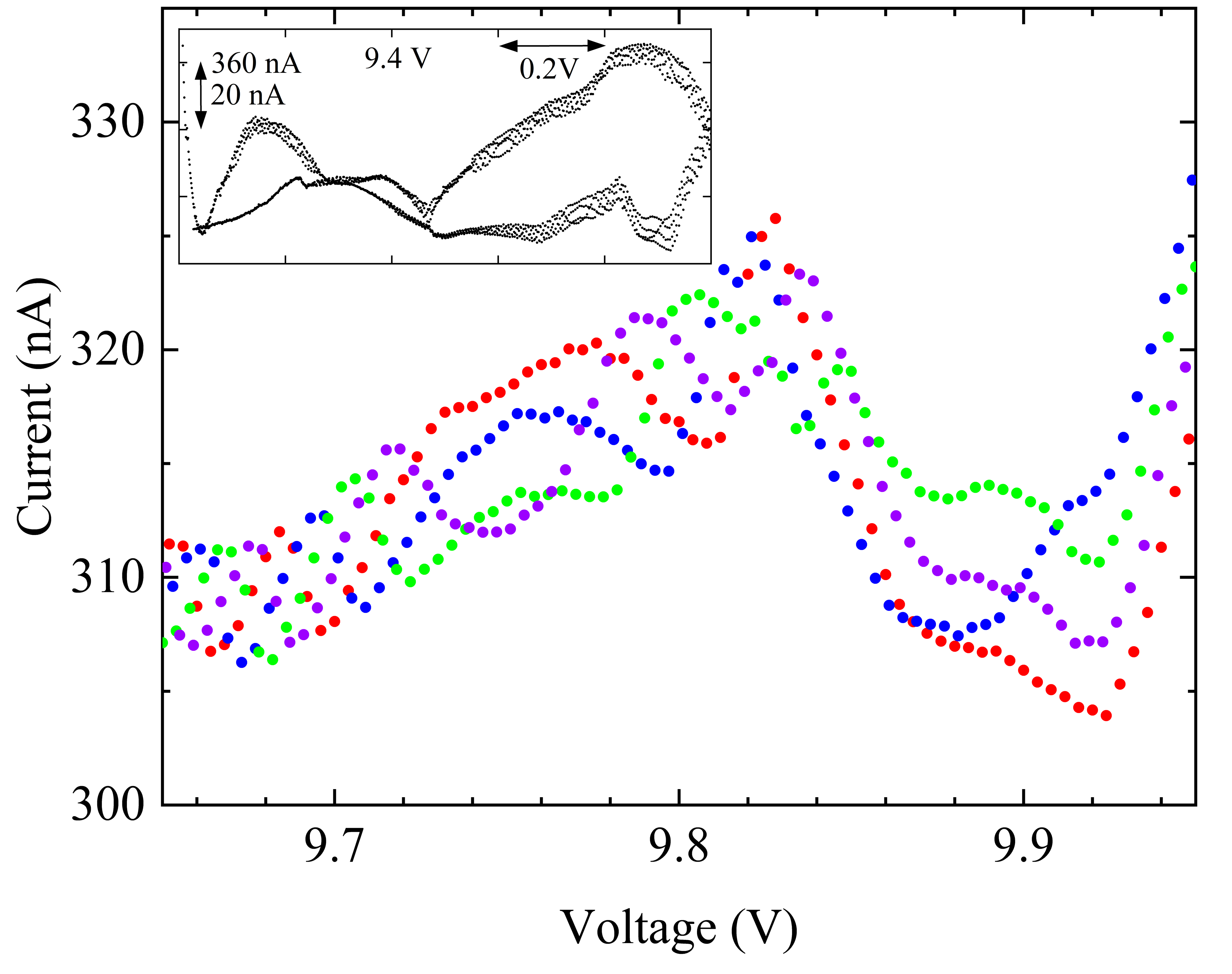}
    \caption{A distinction in four groups of data is visible as colored points in this IV curve from BBA2. Over the entire voltage range the measurement sequence is red, blue, green, purple, red, … and so forth, revealing four state-curves of the molecular system, which take precisely defined turns in carrying the current, triggered by the previous measurement. The inset shows the IV curve over the entire \SIrange{9}{10}{\volt} range. }
    \label{fig:BBA2}
\end{figure}

\begin{figure}[H]
    \centering
    \includegraphics[width=.47
    \textwidth]{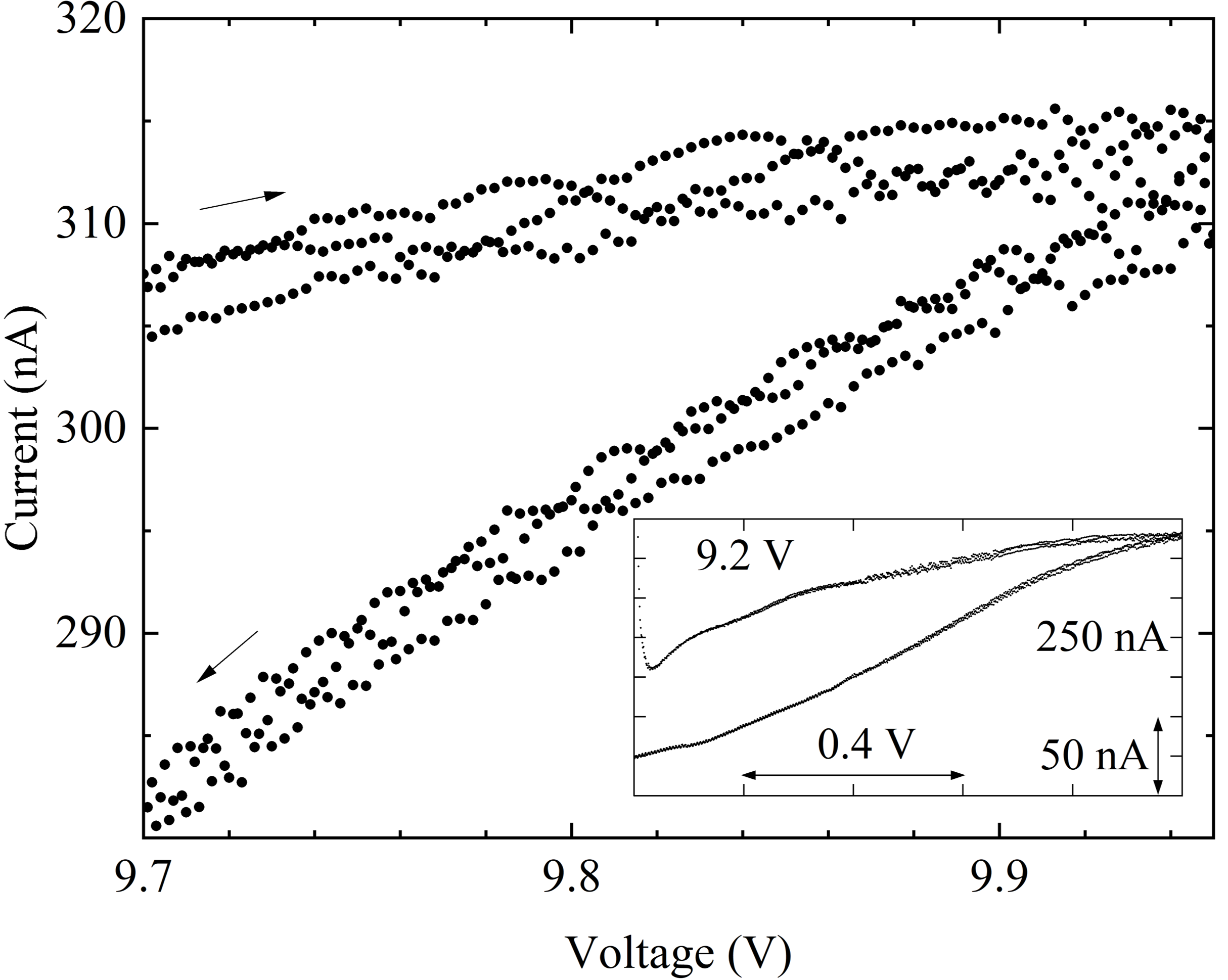}
    \caption{An IV curve from BBA3 revealing a similar pattern as in fig. \ref{fig:IV02}, \ref{fig:IV03a}a and \ref{fig:BBA2}. There are three state-curves involved, in the lower IV curve two of these state-curves seem to interact as a function of voltage. The inset shows the entire IV curve over the \SIrange{9}{10}{\volt} range. }
    \label{fig:BBA3}
\end{figure}

The average current of an IV curve provides a relative indication of the partially wet phase thickness. At the start of most experiments this layer grows, nearing the end it will reduce in thickness. During the start and end of the partially wet phase the average conduction is typically low, in between it is high. Fig. \ref{fig:IV02}, \ref{fig:IV03a}a, \ref{fig:IV04} display an average current of \SI{180}{\nano\ampere}, for fig. \ref{fig:IV01} and \ref{fig:IV05} it is in the \SIrange{30}{50}{\nano\ampere} range.

Fig. \ref{fig:IV01} is recorded at the start of the partially wet phase, the IV curve shows an increasing “noise level” towards higher voltages. The inset reveals that the increased spread of

 measurement points is regular in nature and cannot be attributed to noise. Fig. \ref{fig:IV04} seems to indicate the end of the regular pattern observed in the previous two figures. A noisy IV curve remains, where bits of this IV curve might still show repetitive behavior. In the final IV curve from BBA1, fig. \ref{fig:IV05}, a different pattern appears. This picture shows clear sharp pointed quanta of similar width and height. The inset shows the IV curve over the entire \SIrange{9}{10}{V} range.

% SECTION 3
\section{Theoretical background and discussion}
In recent years, physicists involved in the field of molecular partially wet phase MCB junctions have been increasingly focusing on two specific quantum systems \cite{EU}, \cite{Pollock}. It is a possibility that the quantum effect reported in 2021 \cite{Muller} is an entanglement of two quantum effects. Each of these quantum effects would be destroyed by the absence of the THF partially wet phase. 

The first quantum effect (qe1) \cite{Pollock} relates to a single molecule anchored between two electrodes, lined with an electrically shielding partially wet phase layer. This structure provides for an ideal testbed of open quantum systems. The quantum system couples to the controlled environment inside the Faraday cage which is provided by the floating charge carriers on the partially wet phase. This quantum effect relies on electrical isolation from the environment and thus requires the presence of a THF partially wet phase.

The second quantum effect (qe2) \cite{EU} relates to an increasing or decreasing partially wet phase layer thickness. The microscopic THF layer can increase in thickness due to redeposition of THF vapor from the drying cell on the junction area. The continuously increasing thickness of the partially wet phase will pull the electrodes stronger together, continuously increasing the compressive force on the bridging molecule between the electrodes leading to an increasing average conduction. At a certain point in time the cell is dry and the partially wet phase on the junction area will start to decrease in thickness due to evaporation. We will call this point, which is in between an increasing and a decreasing force on the molecule, the inflection point. The compressive force now turns into a continuously increasing tensile force on the molecule leading to a decreasing average conduction. The continuous increasing or decreasing strain on the molecule leads to quantum interference effects of the various molecular atom positions and/or couplings. This gives rise to the observed reproducing structure in the current as a function of voltage. However, this structure would also show as a function of time at a fixed voltage as the driving force is the varying compressive or tensile force on the bridging molecule resulting from an increasing or decreasing THF partially wet phase layer. At the inflection point the amplitude of the structure becomes zero. This quantum effect requires the presence of a THF partial wet phase, electrically shielding the environment. 

\begin{figure}[H]
    \centering
    \includegraphics[width=.45\textwidth]{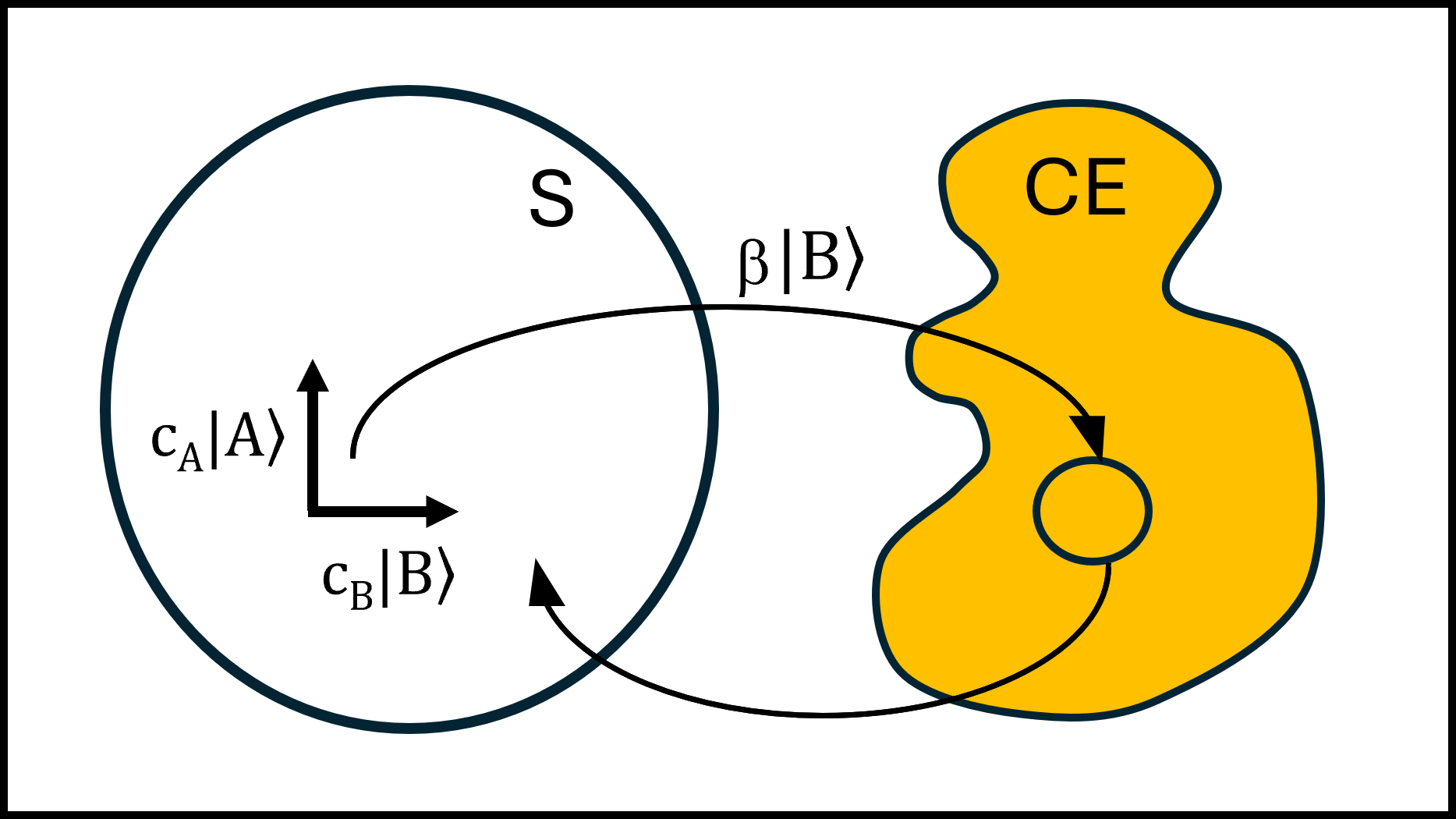}
    \caption{A schematic representation of the quantum system S coupled to the controlled environment CE.
Some $\beta\ket{B}$ probability flowed from S to CE, which is allowed to (partially) backflow to S.}
    \label{fig:nonMarkov1}
\end{figure}

Here we focus on effects which are likely related to qe1, ideally these experiments are executed at the inflection point. In fig. \ref{fig:nonMarkov1} we have indicated a quantum system (S) coupled to a controlled environment (CE) for a non-Markovian system where the information flow from the quantum system to the controlled environment is also allowed to flow back to the quantum system. Translating the open quantum system to reality, the molecule captured between the two Au electrodes provides for the quantum system. The floating charge carriers at the interface of the THF partially wet phase and air have the characteristic of providing for a Faraday cage for the enclosed molecule. Therefore, the outside environment is electrically isolated. This caters for a controlled environment inside the Faraday cage: the THF partially wet phase. The quantum system is coupled to the controlled environment which has the advantage that environmental modes from outside the Faraday cage are unable to have any influence on the delicate S-CE coupled system. Let us assume that the quantum system consists of two states, $\ket{A}$ and $\ket{B}$. The normalized wavefunction is provided by $\ket{\psi}=c_A\ket{A} +c_B\ket{B}$ where $c_A$ and $c_B$ are the probability amplitudes: $c_A^2 + c_B^2 =1$. Over time, there will be a flow of information from S to CE. This can be information with $\ket{A}$ character or information with $\ket{B}$ character or information with both $\ket{A}$ and $\ket{B}$ character. We will show below that the measurement outcome will be alternating no matter what the start conditions are, for now we assume some $\beta \ket{B}$ probability has leaked to CE. It is important to note that the controlled environment together with the increased decoherent quantum system still contain all the initial quantum system information, enabling one to reconstruct the original quantum system. The arrow from CE to S indicates that backflow of (part of) the leaked information is possible. The important question is, what happens when a measurement is performed?

\begin{figure}[H]
    \centering
    \includegraphics[width=.92\textwidth]{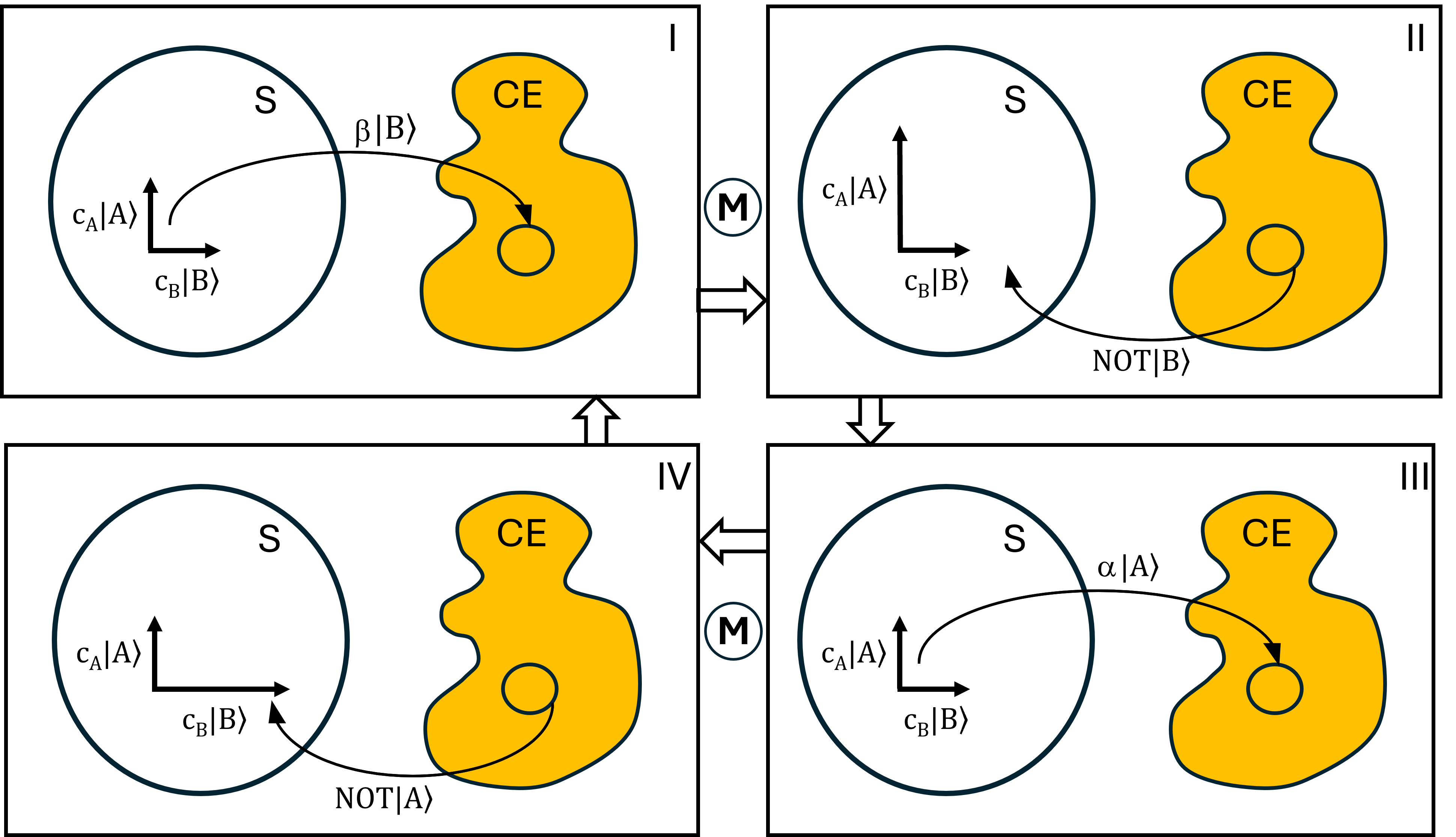}
    \caption{Pictographic tomography of the influence of a measurement on a non-Markovian system. The four indicated situations, I to IV, represent a measurement cycle containing two different measurements indicated by the encircled "M’s". This cycle constitutes the origin of the observed alternating measurements. The probability amplitudes $c_A$ and $c_B$ are situation I-IV specific.}
    \label{fig:nonMarkov2}
\end{figure}

In fig. \ref{fig:nonMarkov2} the influence of a measurement on S is studied in detail. In situation I a certain amount of $\ket{B}$ probability has flown from S to CE. Subsequently a measurement is performed, indicated as the encircled “M” between situation I and II. The measurement outcome will be $\beta\ket{B}$, a measurement acts on CE, not on S. As soon as the measurement is executed the $\beta\ket{B}$ information in CE is gone. It has proliferated through the system and has ended up as a data point in the measurement equipment. Information that is left behind is NOT$\ket{B}$ which still contains $\ket{A}$ character. This information is allowed to flow back to S (situation II) and will bolster $\ket{A}$. As $\ket{A}$ is now the state with the highest probability, preferably $\ket{A}$ will leak to CE (situation III). A subsequent measurement will erase the $\alpha\ket{A}$ information from CE, as this will end up in the measurement equipment. As a result NOT $\ket{A}$, which may still contain $\ket{B}$ character, is allowed to flow back to S. 

\end{multicols}

\begin{multicols}{2}

Resulting in a bolstered $\ket{B}$ vector (situation IV). This will lead to leaking preferably $\ket{B}$ character to CE, ending up in situation I again. Starting with a similar amount of $\ket{A}$ and $\ket{B}$ character being leaked to CE, in situation I, would not have modified the alternating measurements outcome. The important aspect here is that the measurement equipment can only measure $\ket{A}$ or $\ket{B}$ character, removing one and leaving the other. 

The above cycle explains the origin of the alternating measurements in fig. \ref{fig:IV02} and \ref{fig:IV03a}a. In those sections of the IV curve where two state-curves are present, at a specific voltage, the measurement of one state forces the quantum system in the other state which will be the outcome of the next measurement. Does the above also explain the fixed sequence in the case of three or four state-curves (fig. \ref{fig:BBA2} and \ref{fig:BBA3})? It does assuming an equivalent leak rate of information for the three or four states. We can choose any of the three states, $\ket{A}$, $\ket{B}$ or $\ket{C}$ to be the first one measured, we take $\ket{B}$. The next measurement will be $\ket{A}$ or $\ket{C}$ as the $\ket{B}$ character has been effectively removed by the first measurement, we take $\ket{A}$. The third measurement will then be $\ket{C}$ as $\ket{C}$ character has been building up within CE for the longest time. Next measurement will be $\ket{A}$ again for the same reason, and so forth. A similar argumentation holds in the situation of four states, see fig. \ref{fig:BBA2}. The four discerned state-curves in this figure have been colored over the indicated voltage range. The measurement increments amount to 1 mV, within the same color group the points are 4 mV apart. The measurement sequence is red, blue, green, purple, red.. and so forth for the entire indicated voltage range! This implies that the measurement itself determines the next one, it is not an average of a stochastic process, how a measurement is often considered. The four states of the quantum system take well defined turns in carrying the current, driven by the measurements themselves. Various voltage sections of fig. \ref{fig:IV02}, \ref{fig:IV03a}a, \ref{fig:BBA2}, and \ref{fig:BBA3} have been inspected to confirm the above-described behavior. First the number of state-curves is determined at a specific voltage, from that follows the voltage increment for the data points within each state-curve. Next it was checked that the voltage increments between nearest state-curves increased with 1 mV. The state-curve sequence of being measured remained identical throughout the entire voltage range where the states exist.

In fig. \ref{fig:IV02} and \ref{fig:IV03a}a a min-max current bandwidth limits the range of the data points. In these figures as well as in fig. \ref{fig:IV03c}c the min-max current bandwidth oscillates as a function of voltage. The larger the period is (third panel from above in fig. \ref{fig:IV03c}c) the more well defined the minimum current bandwidth, which is called a “node”. In fig. \ref{fig:IV02} the periods of the min-max current bandwidth become smaller towards larger voltages. In fig. \ref{fig:IV03a}a they become smaller towards lower voltages. In fig. \ref{fig:IV03b}b and \ref{fig:IV03c}c towards smaller periods the nodes become less sharp defined. The data points at or near the nodes start to form additional state-curves. Therefore, the generation or deletion of states or state-curves starts at the nodes. In this way the reducing periods play a role in this process. These extra states also seem to be bound by the same min-max current bandwidth as observed in the rest of the IV curve. Also, it is currently unclear if the bandwidth oscillations are a consequence of a minute changing pressure on the bridging molecule, as in qe2, or if there is a direct voltage dependency only.

In fig. \ref{fig:IV05} the partially wet phase is nearing its end, the THF thickness is reducing because of continuous evaporation. The spiked quanta show qe2 character. However, also qe1 character is often present in these quanta. This normally appears at a particular quant in the IV curve as a minute deviation which grows larger every next quant. This can continue over the entire voltage scale, until the original quant shape can no longer be detected. In that case the non-Markovian character is shown by the continuously modifying quanta, see for example fig. 15 in reference \cite{Muller}. In this reference the time between two quanta equals \mbox{0.5 seconds}, which seem unrealistically high for a quantum effect. However, the data show that the quanta are modified on this characteristic timescale. 

Even though in this paper four IV curves, fig. \ref{fig:IV02}, \ref{fig:IV03a}a, \ref{fig:BBA2} and \ref{fig:BBA3}, from three different bending beam assemblies show similar effects, this is in general difficult to capture. After the third IV curve from BBA1, these effects no longer show in fig. \ref{fig:IV04} and \ref{fig:IV05}. The experimental circumstances must be right to reveal the pure qe1 effect. Although the definition of “right” remains vague for now, we do know that it is favorable to record these effects at or close to the inflection point. 

Another aspect is related to the process described in fig. \ref{fig:nonMarkov2}, where the information leakage between S and CE occurs in a certain characteristic time. The reported IV curves in fig. \ref{fig:IV02}, \ref{fig:IV03a}a, \ref{fig:BBA2} and \ref{fig:BBA3} have obviously been using the right time between data points, 45 ms, in combination with the characteristic time provided by the experiment. As stated above, for a quantum effect this seems like an eternity, yet it is not far off with respect to the characteristic time of 0.5 seconds for the modifying quanta. The exact nature of these long characteristic times (\SI{45}{\milli\second}, \SI{0.5}{\second}) is unclear. Yet experimental data show these long timescales in the partially wet phase MCB junction for two different non-Markovian phenomena with respect to qe1, modifying quanta and alternating measurements. Further experiments with varying scan speed are required to gain more insight into the characteristic times involved in the measurement cycle depicted in fig. \ref{fig:nonMarkov2}.

\section{Conclusions}
An IV measurement performs spectroscopy on the current carrying states of a quantum system coupled to a controlled environment. Non-Markovianity, or memory effects are responsible for the observed alternation of measurements or the fixed sequence of current carrying states being measured after one another. The measurement acts as a sink of information within the controlled environment and therewith influencing the next measurement. The partially wet phase MCB junction is adding an experimental method to investigate a very intriguing phenomenon in physics, the measurement problem.

% \usepackage{booktabs}

\bibliographystyle{unsrt}
\bibliography{references}

\end{multicols}

\end{document}